\documentclass[a4paper]{article}
\usepackage{INTERSPEECH2022}

\usepackage{times}
\usepackage{epsfig}
\usepackage{graphicx}
\usepackage{amsmath}
\usepackage{amssymb}

\usepackage{multirow}
\usepackage{bigstrut}
\usepackage{comment}
\usepackage{bm}
\usepackage[ruled,linesnumbered]{algorithm2e}
\usepackage{color}
\usepackage[pagebackref=true,breaklinks=true,colorlinks,bookmarks=false]{hyperref}



\begin{document}

%%%%%%%%% TITLE
% \title{Unified Online Clustering for Speaker Diarization}
\title{Interrelate Training and Searching: A Unified Online Clustering Framework for Speaker Diarization}

% Interrelate Training and Searching: A Unified Online Clustering Framework for Speaker Diarization
% Interrelate Training and Searching for Online Clustering-based Speaker Diarization
% Coupling Training and Searching for Online Clustering-based Speaker Diarization
% joint optimization of training and Searching for Online Clustering-based Speaker Diarization

% interactive unified online clustering and embedding extraction for Speaker Diarization
%

\name{Yifan Chen$^{1,2}$, Yifan Guo$^{1,2}$, Qingxuan Li$^{3}$,  Gaofeng Cheng$^{1,\dagger}$\thanks{$^\dagger$ Corresponding Author.}, Pengyuan Zhang$^{1,2}$, Yonghong Yan$^{1,2}$}
%The maximum number of authors in the author list is twenty. If the number of contributing authors is more than twenty, they should be listed in a footnote or in acknowledgement section, as appropriate.
\address{
  $^1$Key Laboratory of Speech Acoustics and Content Understanding, Institute of Acoustics, Chinese Academy of Sciences, China,
  $^2$University of Chinese Academy of Sciences, China\\
  $^3$Tsinghua University, Beijing, China}

\email{\{chenyifan,guoyifan,chenggaofeng,zhangpengyuan,yanyonghong\}@hccl.ioa.ac.cn, liqx20@mails.tsinghua.edu.cn}

\maketitle

\begin{abstract}

% 先说流式 to improve online performance 
% 不是细节性 是总体性 对比性 点评性 推广性描述

% For online speaker diarization, samples arrive incrementally and the overall distribution of the samples is invisible. 
% And in most existing clustering based methods, the training objective of the embedding extractor is not optimized specially for clustering. 
% To improve online speaker diarization performance, we propose an online unified clustering framework, which provides an interactive manner between embedding extractors and clustering algorithms. 
% Specifically, it consists of two highly coupled parts, i.e. clustering-guided recurrent training (CGRT), and truncated beam searching clustering (TBSC). 
% The CGRT introduces the clustering procedure into the training process of embedding extractors, which could not only provide cluster-aware information for embedding extractor, but also supply crucial parameters for clustering process afterwards. 
% And with these parameters which contain prior information of the metric space, the TBSC tries to find locally optimal combination of embeddings by penalizing probability score of each cluster.
% Based on this, the TBSC could output clustering results in an online fashion with low latency, which could be seen as a fault-tolerant design for online application. 
% With above innovations, our method achieves 14.48 DER with collar 0.25 at 2s latency on the AISHELL-4, while the DER of the agglomerative hierarchical clustering is 14.57.
For online speaker diarization, samples arrive incrementally, and the overall distribution of the samples is invisible. Moreover, in most existing clustering-based methods, the training objective of the embedding extractor is not designed specially for clustering. 
To improve online speaker diarization performance, we propose a unified online clustering framework, which provides an interactive manner between embedding extractors and clustering algorithms. 
Specifically, the framework consists of two highly coupled parts: clustering-guided recurrent training (CGRT) and truncated beam searching clustering (TBSC). 
The CGRT introduces the clustering algorithm into the training process of embedding extractors, which could provide not only cluster-aware information for the embedding extractor, but also crucial parameters for the clustering process afterward. 
And with these parameters, which contain preliminary information of the metric space, the TBSC penalizes the probability score of each cluster, in order to output more accurate clustering results in online fashion with low latency.
With the above innovations, our proposed online clustering system achieves 14.48\% DER with collar 0.25 at 2.5s latency on the AISHELL-4, while the DER of the offline agglomerative hierarchical clustering is 14.57\%.

\end{abstract}

\section{Introduction}

Speaker diarization aims at detecting the speech activities of each person in a conversation. In other words, it is a task of detecting "who speak when". Recent speaker diarization systems are based on two main approaches, i.e. clustering-based approaches and fully supervised approaches~\cite{anguera2012speaker}. 
Fully supervised approaches optimize diarization objects directly in a fully supervised way and could achieve promising performance, especially when speech from different speakers is highly overlapped~\cite{fujita2019end,medennikov2020target}. However, fully supervised approaches need data with accurate timestamp annotations, which are hard to get in many scenarios.
Compared with fully supervised approaches, clustering-based methods generate diarization results in an unsupervised way by conducting clustering. And thus, only speaker recognition data are required to train the clustering-based systems, which are much easier to obtain. 
% And data with accurate timestamp annotations is not needed. 

% However, in most clustering based methods, the procedures of embedding extracting and clustering are separated during both training and inference, which means the training object of the embedding extractors could be sub-optimal for clustering, and the structural information of clusters are not introduced into the embedding extractors either. 
% Furthermore, with little prior information of the metric space of embeddings, it is difficult for clustering algorithms to decide whether an embedding belongs to a new class or an existing one. As a result, the system performance could be degraded severely, especially in the case of online processing, where embeddings arrive incrementally and the overall distribution of these embeddings is invisible.
% %  for online application, since embeddings arrive incrementally and overall distribution of embeddings is invisible, situation gets worse. 
However, in most clustering-based methods, the procedures of embedding extracting and clustering are separated during both training and inference, which means the training objective of the embedding extractors could be sub-optimal for clustering, and the structural information of clusters are not introduced into the embedding extractors either. Furthermore, with little prior information about the metric space of embeddings, it is difficult for clustering algorithms to decide whether an embedding belongs to a new class or an existing one, especially when considering clustering in an online fashion. 
To solve such problems, we propose a unified online clustering framework which not only optimizes embedding extractor towards online clustering objective, but also provides the clustering algorithm with essential information mentioned above. And our basic idea is that it would be easier for the online clustering algorithm to make a decision when a new sample comes if we could know which samples are definitely from the same class and in which condition we should assign a new sample to a new class. 
% Under this basic idea,  

% Amoung fully supervised approaches, UIS-RNN\cite{zhang2019fully} could 

% the training objective of embedding extractor in clustering based systems is classifying different speakers. And embedding clustering is an unsupervised process. Therefore, there exits a gap between training objective and diarization objective.

% To address this problem, fully supervised approaches are introduced to diarization task. These methods optimize diarization objective directly by using timestamp annotations and could achieve better performance on overlapping speech.  

% The contributions of this work could be summarized into three aspects: 
% (1) We propose a unified training and inference framework for online clustering, which provides an interactive manner between embedding extractors and clustering algorithms.

% (2) Under this framework, an embedding extractor training scheme named clustering-guided recurrent training (CGRT) is proposed to introduce the cluster-aware information into the training process of the embedding extractor, and also provide structure information of the embeddings' metric space for the clustering algorithm simultaneously.

% (3) An online clustering method based on the truncated beam search (TBSC) is further introduced, which utilizes the information supplied by the CGRT to penalize probability scores for potential candidates.

The contributions of this work could be summarized into three aspects: 

(1) We propose a unified training and inference framework for online clustering, which provides an interactive manner between embedding extractors and clustering algorithms.

(2) Under this framework, an embedding extractor training scheme named clustering-guided recurrent training (CGRT) is proposed to introduce the cluster-aware information into the training process of the embedding extractor, and also provide structural information of the embeddings' metric space for the clustering algorithm simultaneously.

(3) An online clustering method based on the truncated beam search (TBSC) is further introduced, which utilizes the information provided by the CGRT to penalize probability scores for potential candidates.

% 为啥要这个，因为 online 对一致性的要求更高，看到整体结构，会更好。。。

% 人数不定是如何解决的，VBX 是假设觉得多的人数，top-down，我们来自相反的方向 bottom-up

% conduct clustering and compare the result with ground truths during embedding extractor training process, trying to introduce cluster-aware information for embedding extractor. Meanwhile, the comparing results could also supply prior information of metric space for clustering algorithm as shown in right part. And truncated beam search tries to find locally optimal combination of embeddings by using the parameters to penalize probability score of each route and outputs results in an online way.

\begin{figure*}[htbp]
	\vspace{-4mm}
	\begin{center}
		\begin{minipage}[t]{0.95\linewidth}
			\includegraphics[width=1.0\linewidth]{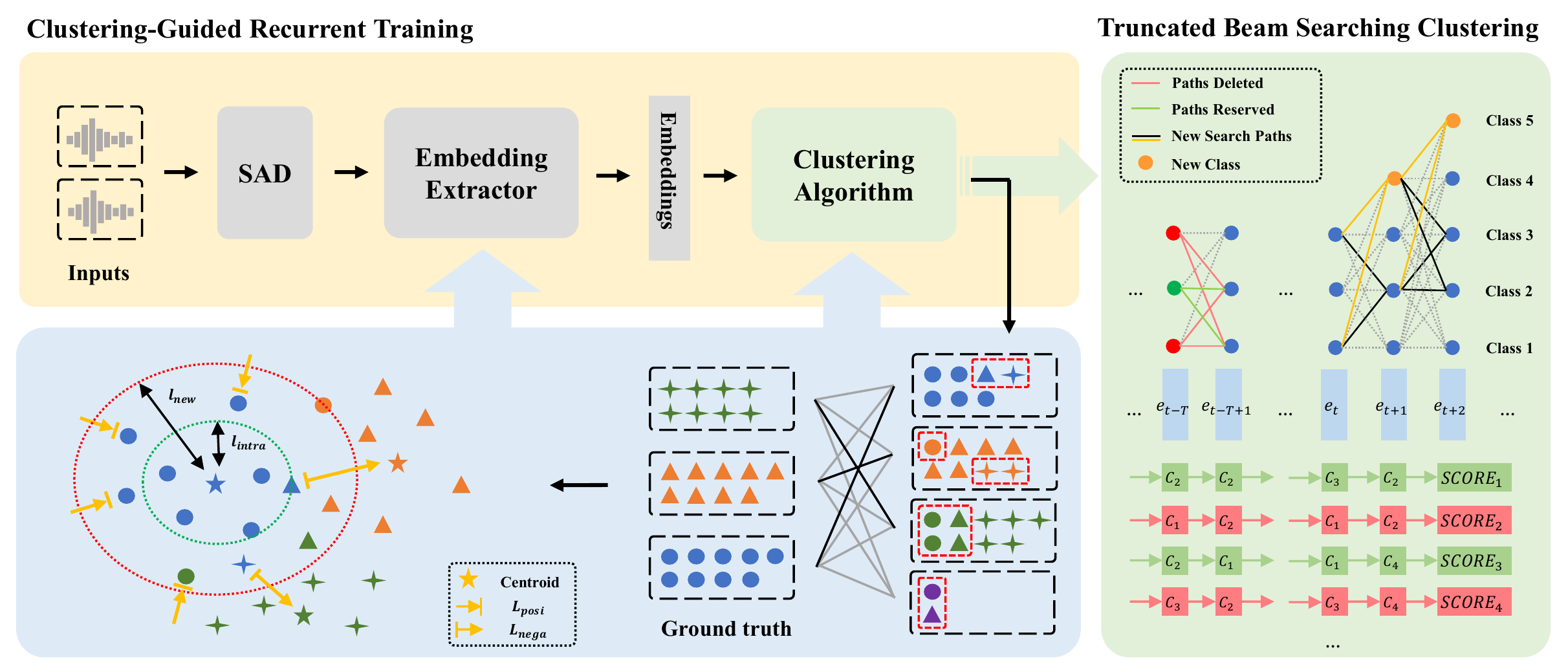}
		\end{minipage}
	\end{center}
	\vspace{-4mm}
	\caption{Structure of the proposed unified online clustering for speaker diarization.
% 	Our online unified clustering framework consist of two tightly coupled parts: an embedding extractor training scheme, i.e. clustering-guided recurrent training and an online clustering algorithm, i.e. truncated beam searching clustering. 
Clustering-guided recurrent training is shown in the left part, where clustering results are compared with the ground truths during the training process of embedding extractor, trying to introduce cluster-aware information for the embedding extractor. Meanwhile, the results of the comparison could also supply prior information of metric space for clustering algorithm as shown in the right part. And the truncated beam searching clustering tries to find locally optimal combination of embeddings by using the parameters to penalize the probability score of each path and outputs results in online fashion. Note that the color of each sample represents its assigned label, while the shape represents the ground truth.}
	\label{image:structure}
\end{figure*}

\section{Related Work}

\subsection{Speaker diarization}

% \subsubsection{Clustering based speaker diarization}
% Clustering based system mainly consists of three steps: speaker activity detection, speaker embedding extracting and clustering. Extracting representations of speakers has been well studied in recent years, and promising results have been achieved~\cite{dehak2010front,variani2014deep,snyder2018x}. 
For a clustering-based system, the typical method is to conduct the hierarchical clustering~\cite{johnson1967hierarchical} or the spectral clustering~\cite{ng2001spectral} directly on the extracted embeddings such as i-vectors, d-vectors, and x-vectors~\cite{dehak2010front,variani2014deep,snyder2018x}. 
However, the direct use of these algorithms does not consider the diarization task's specialty.
Therefore, several sophisticated clustering models are designed. Fox et al.~\cite{fox2007sticky} introduce the Dirichlet processes mixture model into the diarization task. 
Diez et al.~\cite{diez2019analysis,diez2019bayesian,landini2022bayesian} designed a Bayesian model, i.e. VB-HMM, to conduct clustering, whose hidden states correspond to the speaker identities while the observable states correspond to the embeddings.
% The VB-HMM is based on the hypothesis that a conversation could be seen as a Hidden Markov process, whose hidden states correspond to the speaker identities while the observable states correspond to the embeddings.
Besides, graph neural networks are used for clustering by Wang et al.~\cite{wang2020speaker}, where the structural information of the embeddings is utilized. However, all these methods conduct clustering in an offline way. And since they need knowledge about the whole distribution of the samples, it is difficult to extend these methods to online inference.
To alleivate the training-inference mismatch problem in the clustering based framework, fully supervised methods are introduced. UIS-RNN~\cite{zhang2019fully} uses the recurrent neural networks (RNN) to model the transition probabilities among different speakers, which could output diarization results in an online fashion.
% in recent years. 
EEND~\cite{fujita2019end,horiguchi2020end,maiti2021end} discards the use of the speakers' embeddings and directly optimizes the diarization task in an end-to-end manner. Moreover, Xue et al.~\cite{xue2021online} introduce a speaker-tracing buffer in order to solve the across-chunk permutation issue when extending the EEND to online applications. 
% ~\cite{fujita2019
% All these methods treat speaker diarization as a sequence-to-sequence problem, aiming to decode a sequence of speaker labels from an initial sequence of features.
% Another perspective is regarding speaker diarization as a object detection task on time-frequency domain, which needs to regress bounding boxes' position of each person. 
% Another view is to regard speaker diarization as an object detection task in the time-frequency domain, where the goal is to regress the position of each person's bounding box.
Besides, there are other ways to conduct diarization. RPNSD~\cite{huang2020speaker} utilizes the Faster R-CNN~\cite{ren2015faster} to conduct the diarization task, which is a commonly used image segmentation approach. And TS-VAD~\cite{medennikov2020target} compares the embedding of a chunk of audio with the embedding of a target speaker to decide whether the chunk includes the corresponding speaker or not.
All these non-clustering based diarization methods require domain-matched training data with accurate timestamp annotations, which are hard to obtain in some scenarios.

\subsection{Metric embeddings learning}

Metric embeddings learning has developed rapidly in recent years, aiming to map semantically similar samples into similar embeddings and map different ones into discrepant embeddings. 
Weinberger et al.~\cite{weinberger2005distance} propose a loss that could pull the data points from the same class together and push the ones belonging to different classes away from each other. Based on it, and benefiting from the development of deep learning, Triplet loss~\cite{schroff2015facenet} is proposed and is applied successfully on large-scale face recognition. And Hermans et al.~\cite{hermans2017defense} optimize hard samples firstly rather than optimize all samples simultaneously. Besides, in order to enhance the discriminability of embeddings, classification margins are introduced~\cite{wang2018additive,deng2019arcface}. All these methods are optimized for verification or identification task, which may be sub-optimal for online clustering applications. And motivated by these cutting-edge methods, our clustering-guided recurrent training does not only constrain the metric space of embeddings, but also provides 
such information for the truncated beam searching clustering algorithm for clustering afterward.

\subsection{Online clustering algorithm}
Benefiting from the development of machine learning, several works conduct clustering in an online fashion. Typical methods cluster incremental samples in a greedy manner~\cite{charikar2004incremental}. However, in this way, 
% assigning a sample to a new class or merging two classes into one class need thresholds, 
thresholds are required to assign a sample to a new class and to merge two classes into one class,
which limits the 
generalization of algorithms in applications. Instead of using a threshold to judge a new sample belongs to which classes, Liu et al.~\cite{lilt2004online} proposes within-cluster dispersion as a criterion for classes selection. And particle filters are introduced to the online application by Mansinghka et al.~\cite{mansinghka2007aclass}, which use a sequential Monte Carlo scheme to approximate the inference of the Chinese restaurant process mixture model. 
% However, all these online clustering methods suffer from lacking overall distribution of samples and knowing litte prior information of samples' metric space. 
However, all these methods suffer from the lack of the prior information on the samples' overall distribution and metric space.
Therefore, it is still challenging for them to decide whether a new sample belongs to a new class or an existing one.

\section{Our Approach}

% 由于在线聚类无法看到全局信息，会给聚类带来一系列的困难，如新类别的划分。因此对于 embedding 的准确性与一致性的要求更高。为了解决这个问题，我们提出了 聚类一致性约束 的训练方式。约束看到整体结构，会更好。。。
% 并且在此之上，使用截断的 beam search 来进行在线类别的划分。

% However, in most clustering based methods, embedding extracting and clustering are separated during both training and inference, which means embedding extractor training object may be sub-optimal for clustering, and structural information of clusters are not introduced to embedding extractor. Furthermore, with little prior information about the metric space of embeddings, it is difficult for clustering algorithm to assign embeddings into a new class or an existing one. And for online application, since embeddings arrive incrementally, the center of a cluster is varying and the whole distribution of samples is unknown during clustering process. Therefore, situation gets more complicated when considering clustering in an online fashion. Nevertheless, if we could know which samples are definitely from the same class and in which condition we should assign a new sample to a new class, it would be more easier for online clustering algorithm make a decision when a new sample comes. Under this basic idea, we propose an online unified clustering framework which could not only optimize embedding extractor towards online clustering object, but also provide clustering algorithm with these essential information mentioned above. 

% For online application, embeddings arrive incrementally and 

As shown in Fig.~\ref{image:structure}, our unified online clustering framework consists of two highly coupled parts: an embedding extractor training scheme, i.e. clustering-guided recurrent training (CGRT), and an online clustering algorithm, i.e. truncated beam searching clustering (TBSC). Clustering-guided recurrent training aims to optimize embedding extractor towards online clustering object, and simultaneously provide prior information of metric space for the clustering algorithm. TBSC tries to conduct clustering and output results in an online fashion with the assistance of these prior information.
More details about CGRT and TBSC are as follows.

% introduce cluster-aware information for embedding extractor

% clustering-guided recurrent training conducts embeddings extractor training and clustering alternatively, in which clustering results are compared with ground truths, providing hyperparameters for both extractor optimizer and clustering algorithm. And truncated beam searching clustering uses these hyperparameters to conduct clustering, trying to find locally optimal combination of embeddings output results in an online fashion. The more details about CGRT and TBSC as follows. 

% In training stage, the hyperparameters and embeddings extractor are updated recurrently. For inference,  

% to guide the following optimizing process. 

% On the one hand, CGRT constraints metric space of embeddings which not only enhance the discriminability of embedding extractor, but also facilitates the processes online clustering by providing the information of metric space. On the other hand, truncated beam searching clustering (TBSC) uses the information to obtain a

% better clustering result and provide . For inference, the process is as same as the prediction stage in CGRT as shown in left top of Fig.~\ref{image:structure}. 

We will use the following notations to introduce our method. Assume $\mathcal{X}=\{{\bm{x}_i\}_{i=1}^{N}}$ is a set of training samples, and the corresponding speaker labels is $\{{y_i\}_{i=1}^{N}}$. Embedding extractor $\{\mathcal{F}_t\}_{t=1}^{T}$ which is trained after $t$ iterations, maps training samples $\{{\bm{x}_i\}_{i=1}^{N}}$ to embeddings $\{{\bm{e}^t_i\}_{i=1}^{N}}$. The embeddings are clustered into $K$ clusters, $\{(\mathcal{C}_i, \bm{\mu}_i)\}_{i=1}^{K}$, where $\mathcal{C}_i$ represents embedding set, and $\bm{\mu}_i$ is the corresponding center.

% The embeddings are clustered into $K$ clusters, $\{(\mathcal{C}_i, C_i, \bm{\mu}_i)\}_{i=1}^{K}$, where $\mathcal{C}_i$ represents embedding set, and $C_i$ is the corresponding label of the cluster, and $\bm{\mu}_i$ is the center of embedding set.

% Notes that $\{{\bm{e}_i^t\}_{i=1,\dots,N, t=1,\dots,T}}$ represents embeddings obtained by embedding extractor at different iter. 

\subsection{Clustering-Guided Recurrent Training}

In order to optimize embedding extractor towards clustering object and provide practical information for clustering algorithm, it is essential to know which samples will lead to clustering mistakes. Therefore, the basic idea is to conduct clustering and embedding extractor training alternatively in a prediction-correction way. Specifically, in prediction stage, the embedding extractor provides intermediate embeddings to the clustering algorithm and obtains clustering results. In correction stage, clustering results are compared with ground truths, thus providing feedback to embedding extractor and clustering algorithm, i.e. guiding the coming training process of embedding extractor and updating parameters containing information of metric space for the clustering algorithm.

\begin{algorithm}
	\footnotesize
	\caption{Clustering-Guided Recurrent Training.}
	\label{algo:CGRT}
	\KwIn{embedding extractor $\mathcal{F}$, clustering algorithm $\mathcal{H}$, training set $\mathcal{X}$}
	% \ENSURE 
	% \STATE $y \gets 1$
	\For{$t = 1,2,\dots,T$}
	{
	    \textbf{// prediction stage}\\
	    Sampling an subset $\mathcal{X}'_t = \{({\bm{x}_i, y_i)\}_{i=1}^{M}}$ \\
	    Calculate embeddings $\{{\bm{e}_i^{t}\}_{i=1}^{M}}$ using $\mathcal{F}_{t-1}$\\
	    \textbf{// correction stage} \\
	    Cluster $\{{\bm{e}_i^{t}\}_{i=1}^{M}}$ and get $l_{intra}$, $l_{new}$\\
% 	\STATE find maximum weight matching between clusters and ground truths 
	    Calculate $\mathcal{L}_{posi}$, $\mathcal{L}_{nega}$ with $l_{intra}$, $l_{new}$, $\{{\bm{e}_i^{t}\}_{i=1}^{M}}$ \\
	    Conduct optimization with $\mathcal{L}_{posi}$, $\mathcal{L}_{nega}$ to get $\mathcal{F}_{t}$\\
	}

\end{algorithm}

We now describe more specific steps of our proposed training scheme. As shown in algorithm~\ref{algo:CGRT}, CGRT consists of the following steps. 
% Firstly, embedding extractor is trained on speaker classification task for several iterations. 
% And the forthcoming stage is to conduct prediction-correction.
Suppose that we have a pre-trained feature extractor $\mathcal{F}_{0}$.
In prediction stage, a subset $\mathcal {X}'_t \subset \mathcal {X}$ is sampling from the whole training set randomly. And embedding extractor $\mathcal{F}_{t-1}$ provides intermediate embedding set $\{{\bm{e}_i^t\}}$ corresponding to $\mathcal {X}'_t$. Then, we conduct clustering on these embeddings and obtain clustering results $\{\mathcal{C}_i\}$. 
% Note that we adopt spectral clustering algorithm in at the beginning of iterations, and use online clustering algorithm\ref{section:clustering} afterwards. 

In correction stage, we need to distinguish "positive" embeddings and "negative" embeddings in a cluster. To this end, we firstly find the perfect matching between obtained clusters and ground truths and assign each embedding a "positive" or "negative" label. 
Assume $K$ is the number of the clusters, $\{\mathcal{Y}_j\}_{j=1}^K$ are the speaker label sets generated by the clustering, and $\{\mathcal{G}_i\}_{i=1}^L$ are sets that consist of real speaker labels of samples (ground truths). $L$ is the real number of speakers.
% These speaker labels could be divided into $\{\mathcal{Y}_j\}$ spontaneously according to themselves. 
% The weight of bipartite graph composed by $\{\mathcal{C}_i\}$ and $\{\mathcal{Y}_j\}$ are defined by:
% \begin{equation}\label{equ::weight}
% 	W_{i,j} = \frac{|\mathcal{C}_i\cap\mathcal{Y}_j|}{|\mathcal{C}_i\cup\mathcal{Y}_j|}*|\mathcal{Y}_j|
% \end{equation}
% Using KM algorithm, an perfect matching could be obtained. 
Then we can get a perfect matching using the KM algorithm with a bipartite graph defined by $W_{i,j} = \frac{|\mathcal{G}_i\cap\mathcal{Y}_j|}{|\mathcal{G}_i\cup\mathcal{Y}_j|}*|\mathcal{Y}_j|$, and thus each embedding $\bm{e}_i$ could be assigned to the "positive" set $\mathcal{E}_\text{pos}$ or the "negative" set $\mathcal{E}_\text{neg}$.

Next, we calculate two hyperparameters that contain practical information for both clustering algorithm and embedding extractor optimizing algorithm, i.e., $l_{intra}$ and $l_{new}$:
% \begin{equation}\label{equ::threshold_same}
% 	l_{intra} = \underset{\forall y_i \neq y_j }{\min} \{d(\bm{e}_i, \bm{e}_j)\}
% \end{equation}
% \begin{equation}\label{equ::threshold_different}
% 	l_{new} = \underset{\forall y_i = y_j}{\max} \{d(\bm{e}_i, \bm{e}_j)\}
% \end{equation}
\begin{equation}\label{equ::threshold_same}
	l_{intra} = \underset{\bm{e}_{i}\ \in\ \mathcal{E}_\text{neg}}{\min} \{d(\bm{\mu}_{false}, \bm{e}_i)\}
\end{equation}
\begin{equation}\label{equ::threshold_different}
	l_{new} = \underset{\bm{e}_{i}\ \in\ \mathcal{E}_\text{pos}}{\max} \{d(\bm{\mu}_{true}, \bm{e}_i)\}
\end{equation}
where $d(\bm{x}, \bm{y})$ means the cosine distance between two vectors. The meaning of $l_{intra}$ is the minimum distance of negative embeddings from centers of false classes. And below this threshold, the embeddings should belongs to the existing class. The meaning of $l_{new}$ is the maximum distance of embeddings from centers of true classes. The embeddings should come from a new class when all distances from existing centers of classes are larger than it. 
% This threshold could be adopted to judge whether a sample belongs to a new class or an existing one. 
% Notes that for online clustering, a new sample will

Using these hyperparameters that include information about metric space, the embedding extractor could be optimized with the following two losses:
% In order to optimize difficult samples in training set, we use a loss guided by hyperparameter $l_{intra}$ and $l_{new}$.
% \begin{equation}\label{equ::loss_true_examle}
% 		\mathcal{L}_{posi} = \frac{1}{\mathcal{N}} \sum_{\bm{e}_i,\bm{e}_j\ \in\ \mathcal{E}_\text{pos}} max\{ d(\bm{e}_i,\bm{e}_j) - l_{intra}, 0\}
% \end{equation}
\begin{equation}\label{equ::loss_true_examle}
		\mathcal{L}_{posi} = \frac{1}{\mathcal{N}} \sum_{i,j,y_i=y_j} max\{ d(\bm{e}_i,\bm{e}_j) - l_{intra}, 0\}
\end{equation}
% and for negative samples, we have
\begin{equation}\label{equ::loss_false_example}
	\mathcal{L}_{nega} = \frac{1}{\mathcal{N}} \sum_{\bm{e}_{i}\ \in\ \mathcal{E}_\text{neg}} max\{d(\bm{\mu}_{true}, \bm{e}_i) - d(\bm{\mu}_{false}, \bm{e}_i)+ l_{new} ,0\}
\end{equation}
where $\mathcal{L}_{posi}$ aims at pulling samples which belong to the same class together. And $\mathcal{L}_{nega}$ targets at pushing negative data points from the false center $\bm{\mu}_{false}$ further away, and pulling them towards the center $\bm{\mu}_{true}$ they belong to.

\subsection{Truncated Beam Searching Clustering}
\label{section:clustering}

For online clustering applications, it is not easy to decide whether a new sample belongs to a new cluster or an existing one since samples arrive incrementally. To address this problem, on the one hand, clustering-guided recurrent training is proposed above, which could not only constrain embeddings' distribution in metric space, but also provides hyperparameters containing information about metric space. On the other hand, a clustering method using a heuristic search algorithm is proposed as a fault-tolerant design for online applications. 

The more specific steps of truncated beam searching clustering are as follows. It requires two essential hyperparameters: beam size $B$ and latency $T_0$. The beam size determines the number of search paths preserved during search process. And latency $T_0$ determines truncated length of each search route, thus determining the latency between inputs and outputs. When a new sample arrives, the probabilities of all possible classes are calculated. The details are shown in algorithm~\ref{algo:beam_search}.

% \begin{algorithm}
% 	\caption{\textbf{Truncated Beam Searching Clustering.}}
% 	\label{algo:beam_search}
% 	\begin{algorithmic}
% % 	\hangafter 1
%     % \hangindent 1.2em{
% 	\REQUIRE Beam size $B$, Latency $T_0$, Embedding sequence $\{\bm{e}_i\}$,$i\in[1,T]$, $l_{intra}$,$l_{new}$
% 	\STATE Initialize a max-queue (paths set) $Q=\emptyset$
% 	% \INI
% 	\FOR{$t = 1,2,\dots,T$}
% 	\STATE Calculate scores of assigning $\bm{e}_t$ to a new class ($s(C_{new})$) \\ \quad and existing classes($s(C_{i})$).  
% 	\STATE Use these scores to extend paths to the candidate classes of \\ \quad $\bm{e}_t$ respectively.
% 	\STATE Output the class label at $t-T_0$ corresponding to the current \\ \quad max-score paths as $result_{t-T_0}$.
% 	\STATE Delete paths not containing $result_{t-T_0}$.
% 	\STATE Truncate the history of reserved paths before $t-T_0$.
% 	\IF{$|Q| > B$}
% 	\STATE Shrink $Q$ by score.
% 	\ENDIF
% 	\ENDFOR
% % 	}
% 	\end{algorithmic}
% \end{algorithm}

\begin{algorithm}
\footnotesize
	\caption{Truncated Beam Searching Clustering.}
	\label{algo:beam_search}
	
% 	\hangafter 1
    % \hangindent 1.2em{
	\KwIn{Beam size $B$, Latency $T_0$, Embedding sequence $\{\bm{e}_i\}$,$i\in[1,T]$, $l_{intra}$,$l_{new}$}
	Initialize a max-queue (paths set) $Q=\emptyset$\\
	% \INI
	\For{$t = 1,2,\dots,T$}
	{
    	Calculate scores of assigning $\bm{e}_t$ to a new class: $s(C_{new})$ and existing classes: $s(C_{i})$.\\  
    	Use these scores to extend paths to the candidate classes of  $\bm{e}_t$ respectively.\\
    	Output the class label at $t-T_0$ corresponding to the current max-score paths as $result_{t-T_0}$.\\
    	Delete paths not containing $result_{t-T_0}$.\\
    	Truncate the history of reserved paths before $t-T_0$.\\
    	\If{$|Q| > B$}
    	{
    	    Shrink $Q$ by score.\\
    	}
	}
\end{algorithm}

In the score calculating step in algorithm~\ref{algo:beam_search}, we add some constraints to our search path. If all distances between centers of clusters and the new sample are larger than $l_{new}$, there is a great possibility that the new sample comes from a new cluster. Therefore, a high score will be assigned to the path which leads to a new cluster. And if the distance between a new sample and the center of a cluster is smaller than $l_{intra}$, it would be potential that the new sample belongs to this cluster. Besides, we penalize the continuity of label sequence based on the fact that, in most cases, the current frame and the previous frame belongs the same class.
\begin{equation}
	\label{equ::score_new}
	s\left(C_{new}\right) = 
	\left\{
		\begin{array}{ll}
			S_0, & \min_{j}\left\{d\left(\bm{e}_t, \bm{\mu}_j\right)\right\} \geq l_{new}  \\
			f(\min_{j}\{d(\bm{e}_t, \bm{\mu}_j)\}), & \min_{j}\{d(\bm{e}_t, \bm{\mu}_j)\} < l_{new}
		\end{array}
	\right.
\end{equation}
\begin{equation}
	\label{equ::score_cls}
	s\left(C_i\right) = \left\{
		\begin{array}{ll}
			S_1 + \lambda\cdot\delta_{C^{t-1}, C_i}, & d(\bm{e}_t, \bm{\mu}_i) \leq l_{intra} \\
			g(d(\bm{e}_t, \bm{\mu}_i)) + \lambda\cdot\delta_{C^{t-1}, C_i}, & d(\bm{e}_t, \bm{\mu}_i) > l_{intra}
		\end{array}
		\right.
\end{equation}
Note that the parameters of truncated beam searching clustering for inference could use the last ones in the training stage or use the ones generated on the eval set if available.

% \begin{algorithm}
% 	\caption{\textbf{truncated beam searching clustering.}}
% 	\label{algo:beam_search}
% 	\begin{algorithmic} 
% 	\REQUIRE beam size $k$, data sequence $\textbf{x}_i$,$i\in[1,N]$,
% 	\ENSURE $y = x^n$
% 	\STATE $y \gets 1$
% 	\IF{$n < 1$}
% 	\FOR{i }
% 	\STATE $y \gets 1$
% 	\ENDFOR
% 	\STATE $X \gets 1 / n + 2$
% 	\STATE $N \gets -n$
% 	\ELSE
% 	\STATE $X \gets x$
% 	\STATE $N \gets n$
% 	\ENDIF
% 	\WHILE{$N \neq 0$}
% 	\IF{$N$ is even}
% 	\STATE $X \gets X \times X$
% 	\STATE $N \gets N / 2$
% 	\ELSE[$N$ is odd]
% 	\STATE $y \gets y \times X$
% 	\STATE $N \gets N - 1$
% 	\ENDIF
% 	\ENDWHILE
% 	\end{algorithmic}
% \end{algorithm}

% truncated beam search
% involves continue

\section{Experiments}
\subsection{Datasets}

% We use two groups of open source datasets to validate our proposed method on two languages, i.e. VoxCeleb1\&2~\cite{nagrani2017voxceleb,chung2018voxceleb2} with AMI~\cite{renals2007recognition} for English and CN-Celeb1\&2~\cite{fan2020cn,li2022cn} with AISHELL-4~\cite{fu2021aishell} for Chinese. VoxCeleb1\&2~\cite{nagrani2017voxceleb,chung2018voxceleb2} are two widely used large-scale speaker recognition corpora, which are collected from multi-media and contain 1251 and 6112 speakers correspondingly. Besides, the audio length are 352 and 2442 hours respectively. CN-Celeb1\&2~\cite{fan2020cn,li2022cn} are two large-scale Chinese speaker recognition corpora. CN-Celeb1~\cite{fan2020cn} consists of 274 hours audio from 1000 speakers, while CN-Celeb2~\cite{li2022cn} contains 1090 hours audio from 3000 speakers. We use these datasets to train embedding extractor. For diarization task, AMI~\cite{renals2007recognition} and AISHELL-4~\cite{fu2021aishell} are adopted. They are two open source meeting speech recognition datasets containing accurate timestamp. And the number of speakers in each session of AMI~\cite{renals2007recognition} is 4, while AISHELL-4~\cite{fu2021aishell} is 4 to 8. We use the two datasets to test diarization performance. Note that we only use the test set of AISHELL-4~\cite{fu2021aishell}.

We use open-source datasets to validate our proposed method, i.e. CN-Celeb1\&2~\cite{fan2020cn,li2022cn} and AISHELL-4~\cite{fu2021aishell}. CN-Celeb1\&2~\cite{fan2020cn,li2022cn} are two large-scale Chinese speaker recognition corpora. CN-Celeb1~\cite{fan2020cn} consists of 274 hours audio from 1000 speakers, while CN-Celeb2~\cite{li2022cn} contains 1090 hours audio from 2000 speakers. We use these datasets to train the embedding extractor. For speaker diarization task, AISHELL-4~\cite{fu2021aishell} dataset is adopted, which is an open-source meeting speech recognition datasets containing accurate timestamps. And the number of speakers in each session is 4 to 8. Note that we only use the test set of AISHELL-4~\cite{fu2021aishell}.

\subsection{Implementation details}

For implementation details, an energy-based voice activity detector (VAD) implemented by kaldi~\cite{povey2011kaldi} is adopted to remove silence. ResNet-50 with two fully connected layers is employed to conduct the speaker classification task, and the raw waveform is split into 4s pieces as input. During clustering-guided recurrent training, in each circulation, audios are sampling from 4-8 speakers. And we firstly adopt spectral clustering to provide initial results. Then, TBSC with beam size 1 is employed in order to speed up the CGRT process. And we adopt cosine distance to conduct optimizing procedures. For the clustering task, embeddings are extracted every 100ms, and the length of the windows is 1s. In equation~\ref{equ::score_new} and~\ref{equ::score_cls}, we adopt $f(x)= log(x)$, $g(x)= log(1-x)$, $S_0 = S_1 = 0$.

% 40 dimensional mel frequency cepstral coefficient (MFCC) with 25 ms frame length and 10 ms stride as input feature to detect the speech activity. ResNet-101 with two fully connected layers is employed to conduct speaker classification task with 64 dimensional filterbank features extracted every 10 ms with 25 ms window, and additive margin softmax \cite{wang2018additive} is used to get more distinct decision boundary. Raw waveform is split every 4s (400 dimension) to form ResNet input. We train the speaker embedding network using stochastic gradient descent (SGD) optimizer with $0.9$ momentum factor and $0.0001$ L2 regularization factor. 

\subsection{Results}

\begin{table}[htbp]
    \footnotesize
	\label{sd_result}
	\caption{Speaker diarization result on AISHELL-4.}
	\vspace{-3mm}
	\begin{center}
		\begin{tabular}{cccc}
		    \hline
			\multirow{2}{*}{Mode} & \multirow{2}{*}{Methods} & \multicolumn{2}{c}{DER} \bigstrut[t]\\
			\cline{3-4}
			& & collar 0.25 & collar 0   \bigstrut\\
			\hline
			% EEND    & 5.52 & 17.48 & 45.66  \bigstrut[t]\\
			% \hline
			\multirow{3}{*}{Offline}& X-Vector + AHC 							& 14.57 & 23.01  \bigstrut[t]\\
			& X-Vector + SC				 			& 14.31 & 22.53  \\
			& X-Vector + VB-HMM~\cite{diez2019bayesian}	& 11.10 & 16.76  \bigstrut[b]\\
			\hline
			\multirow{2}{*}{Online}& X-Vector + LFC				 			& 17.66 & 23.21 \bigstrut[t] \\
			& X-Vector + DBC~\cite{lilt2004online}	& 17.01 & 22.84 \bigstrut[b]\\
			\hline
			\multirow{2}{*}{Online}& X-Vector + CGRT + LFC				 	& 16.41 & 22.13  \bigstrut[t] \\
			& X-Vector + CGRT + TBSC		 			& 14.48 & 21.01  \bigstrut[b]\\
			\hline
		\end{tabular}
	\end{center}
	\vspace{-4mm}
\end{table}

We carry out experiments to validate the effectiveness of the proposed framework. Firstly, we use X-Vector with PLDA similarity measurement and agglomerative hierarchical clustering (AHC) as an offline baseline. And x-vector with spectral clustering (SC) is also adopted to make a comparison. Secondly, in order to validate our proposed method in the online setting, online clustering methods, i.e. leader-follower clustering (LFC) and dispersion-based clustering (DBC)~\cite{lilt2004online} are also compared. Leader-follower clustering compares a new embedding with all clusters' centroid and assign it to the most similar one or making a new clutter, which could be seen as the following degenerate situation of TBSC: beam size $B=1$, latency $T=0$, and $l_{new}, l_{intra}$ are not used. The result suggests that with the help of CGRT and TBSC, our online framework could get close to offline clustering. And our approach could outperform previous online clustering methods.

\subsection{Analysis of parameters}

\begin{figure}[htbp]
	\vspace{-4mm}
	\begin{center}
		\begin{minipage}[t]{1.0\linewidth}
			\includegraphics[width=1.0\linewidth]{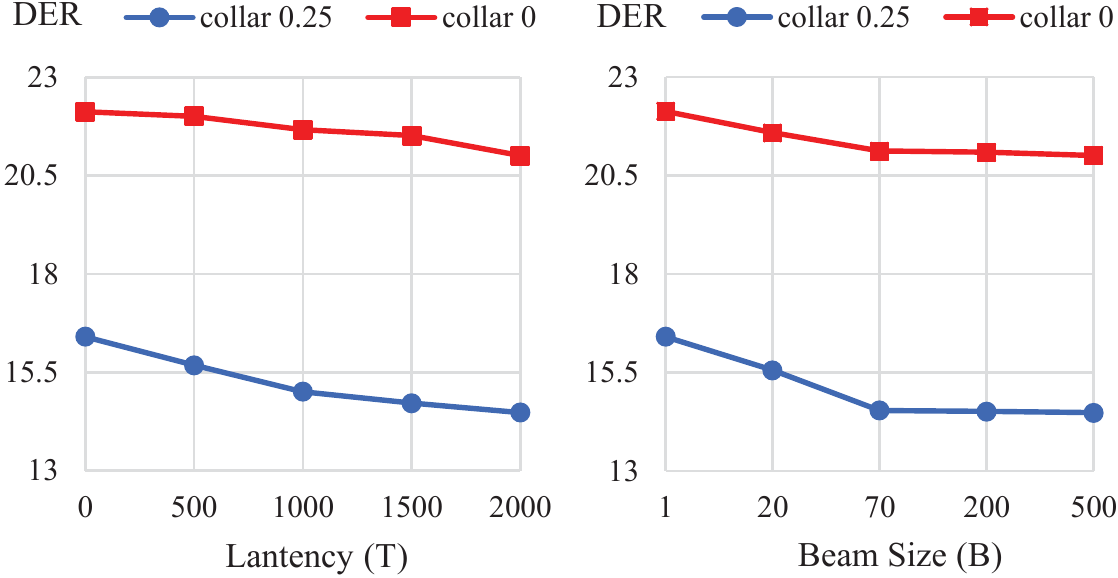}
		\end{minipage}
	\end{center}
	\vspace{-4mm}
	\caption{Impact of the beam size $B$ and latency $T$ in TBSC.}
	\label{image:tradeoff}
\end{figure}

To analyze the effectiveness of our system, we conduct more experiments on the AISHELL-4~\cite{fu2021aishell} dataset. Latency $T_0$ and Beam size $B$ in TBSC are adjusted to test performance. The results show that 
the tradeoff between latency and accuracy exits. And with more memory consumption and latency, TBSC could provide more accurate clustering results. %, which could be seen as a fault-tolerant design for online applications. 

\section{Conclusion}

In this paper, we propose a unified training and inference framework for online clustering. The proposed framework consists of clustering-guided recurrent training (CGRT) and truncated beam searching clustering (TBSC), which provides an interactive manner between embedding extractor and clustering algorithm. Since CGRT introduces the cluster-aware information into the embedding extractor training process, and provides prior information about embeddings' metric space to the clustering algorithm simultaneously, TBSC could conduct online clustering with the knowledge of the coming samples in a matched way. 
With a robust embedding extractor that could initially extract accurate x-vectors, our method achieves competitive results on the diarization task in an online fashion.

\section{Acknowledgment}
This work was partially supported by the Youth Innovation Promotion Association, Chinese Academy of Sciences and the Frontier Exploration Project Independently Deployed by Institute of Acoustics, Chinese Academy of Sciences under Grant QYTS202011.

\bibliographystyle{IEEEtran}
\bibliography{reference}

\end{document}